\documentclass[prl,10pt,twocolumn,notitlepage,aps,floatfix]{revtex4-1}
\usepackage[utf8]{inputenc}
\usepackage{amsmath,amssymb,bm,comment,graphicx}
\usepackage[colorlinks=true]{hyperref}
\newcommand{\D}[0]{\bar{\mathbf{D}}}
\newcommand{\bD}[0]{\bar{\mathbf{D}}}
\newcommand{\change}[1]{{ #1}}

\begin{document}
\title{Compressing and forecasting atomic material simulations with descriptors}
\author{Thomas. D. Swinburne}
\email{thomas.swinburne@cnrs.fr}
\affiliation{Aix-Marseille Universit\'{e}, CNRS,
CINaM UMR 7325, Campus de Luminy, 13288 Marseille, France}
\date{\today}
\begin{abstract}
    Atomic simulations of material microstructure require significant resources 
    to generate, store and analyze. 
    Here, atomic descriptor functions are proposed as a general latent space to 
    compress atomic microstructure, ideal for use in large-scale simulations. 
    Descriptors can regress a broad range of properties, including 
    character-dependent dislocation densities, stress states 
    or radial distribution functions. A vector autoregressive model can 
    generate trajectories over yield points, resample from new initial conditions
    and forecast trajectory futures. 
    A forecast confidence, essential for practical application, is derived by 
    propagating forecasts through the Mahalanobis outlier 
    distance, providing a powerful tool to assess coarse-grained models. 
    Application to nanoparticles and yielding of dislocation
    networks confirms low uncertainty forecasts are accurate 
    and resampling allows for the propagation of smooth microstructure distributions.
    Yielding is associated with a collapse in the 
    intrinsic dimension of the descriptor manifold, which is 
    discussed in relation to the yield surface. 
\end{abstract}

\maketitle
Materials evolve via complex, non-intuitive atomic
mechanisms spanning a wide range of time and length scales\cite{wales_energy_2003,perez2009}.
\change{Atomic simulations (MD) with empirical force fields offer exceptional insight, but although spatial decomposition schemes give excellent (weak) parallel scaling with system size\cite{LAMMPS}, serial time integration limits trajectory duration, irrespective of available processors\cite{perez2015}. 
The ubiquity yet high cost of MD means development of predictive techniques 
to coarse-grain (CG) in space or time is an active research area
\cite{laio2002,perez2009,mardt2018vampnets,van2020roadmap,bonati2021deep,swinburne2022reaction}.}
Material microstructure requires large system sizes, necessitating efficient and scalable
CG techniques. Whilst many structural analysis 
tools exist\cite{honeycutt1987, kelchner1998, ackland2006,lazar2012,lazar2015,stukowski2012automated}, 
none provide generic \textit{compression} of atomic data with a clear metric 
for similarity or diversity, nor is it clear 
\textit{a priori} how to select CG properties,
leading to massive storage requirements at scale\cite{wu2018scalable,zepeda2017probing}.
A further challenge is that simulations of materials are typically non-equilibrium and
exhibit, in part due to timescale limitations, partially disordered structures with a dense 
kinetic spectrum and an unknown steady state, often with external driving\cite{setyawan2015displacement,priezjev2018molecular,mason2020observation}.
To harness modern parallel computers there is thus a recognized need to
\textit{resample} sparse simulation data and to \textit{forecast} simulation futures, both 
for physical insight and to maximize the information yield of additional computational effort\cite{perez2015,swinburne2020automated,garmon2022resource,schaarschmidt2022workflow,zhu2021fully,andrews2022forecasting}.\\
However, the complexities of material deformation 
limit the applicability of current CG and acceleration schemes, 
which require identification of a clear timescale separation\cite{lebris2012} 
to allow parallel time accumulation\cite{voter1998,dimer,voter2000,perez2009,lebris2012,perez2015,swinburne2020automated,swinburne2022reaction} 
or the design of low rank (typically 1-4) collective variables (CV)
which can be used to bias dynamics
\cite{voter1997,neb,laio2002,darve2008,lelievre2010free,bonati2021deep,swinburne2018}. 
Despite many recent advances\cite{rogal2021reaction,bonati2021deep,baima2022} 
general CVs for microstructure remain elusive\cite{swinburne2018,baima2022}, 
instead requiring specialized simulation setups with only a few active mechanisms
such as nucleation\cite{bonati2021deep} or the migration of isolated defects\cite{rogal2021reaction,swinburne2018}.
Exploring \textit{unseen} regions of configuration space 
is known to be uncontrolled as low rank CVs may not remain descriptive\cite{bussi2020using}.
These issues extend to the powerful post-mortem analysis tools\cite{noe2013variational,mardt2018vampnets,huang2017,huang2018direct,
xie2019graph,Rottler2022,mardt2018vampnets,xie2019graph}, which learn collective variables that obey a discrete state Markov model in order to identify kinetically important configurations with implied transition timescales. 
\change{Whilst all-atom\cite{wang2019coarse,klein2023timewarp,han2021artificial,leimkuhler2013stochastic} or coarse-beaded\cite{fu2022simulate} generative models may provide a route for accelerated time-stepping, 
they are currently only competitive to direct time integration for fairly small equilibrium systems 
with a static or slowly varying bonding topology and so cannot be applied 
to large-scale simulations of material deformation where a highly transient, 
heterogeneous atomic connectivity is fundamental.}
%
%
\\
In this contribution, atomic descriptor functions
\cite{Thompson_snap_2015,Shapeev_MTP,goryaeva2021,allen2021atomic,lysogorskiy2021performant}
are proposed as an efficient, general and uncertainty-aware coarse-graining approach, mapping
atomic positions ${\bf X}\in\mathbb{R}^{N\times3}$ to a global vector $\bar{\bf D}\in\mathbb{R}^{\sim100}$, Eq. \ref{eq:mm}. 
\change{The main results are that
1) Descriptors can classify and regress a remarkable range of microstructural properties
(see figures) and permit a data-driven model extrapolation measure\cite{mahalanobis1936generalized}, transferring advances in active learning
\cite{podryabinkin2017active,bernstein2019novo,goryaeva2020dist_score} to atomic CG. This generality 
means CG targets need not be specified \textit{a priori}, giving huge compression in storage and efficiencies in analysis at scale. 
2) Descriptor trajectories can be efficiently resampled and forecasted via a
vector autoregressive (VAR) model \cite{lutkepohl2005new}, with, crucially, a robust 
forecast \textit{uncertainty} derived from the descriptor outlier measure (\ref{tdnmd}).
This allows rapid assessment of when forecasts can be trusted or when additional training is needed, 
essential for practical usage but typically missing in existing schemes. 
The approach is applied to analyze and forecast systems essentially untreatable with existing methods, 
the annealing of large nanoparticles and dislocation yielding under cyclic shear 
and uniaxial tension\cite{zepeda2017probing}. Yielding is identified with a collapse in 
the intrinsic dimension\cite{dadapy} of the descriptor manifold.}

\begin{figure}
    \centering
    \includegraphics[width=\columnwidth]{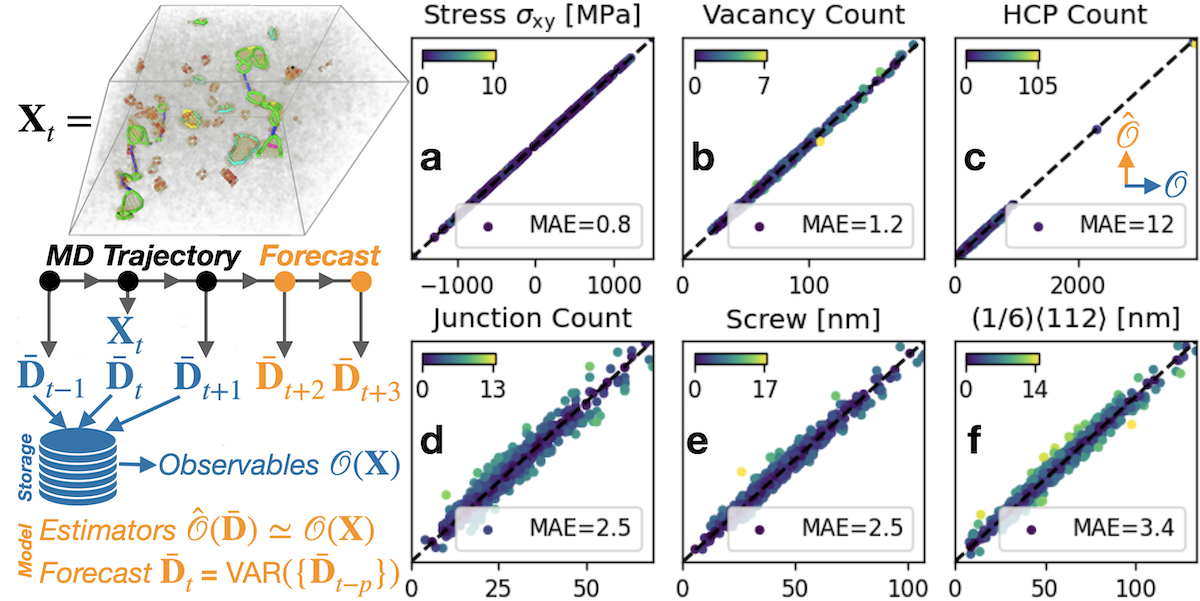}
    \caption{Coarse graining of dislocation networks in Al under cyclic shear, 
    detailed in the SM\cite{SM}.
    Left: Global descriptor vectors $\bar{\bf D}$ (\ref{eq:mm}) are stored every 1-10ps and positions 
    ${\bf X}$ every 100-500ps. $\{{\bf X},\bar{\bf D}\}$ data is
    used to train estimators $\hat{\mathcal{O}}(\bar{\bf D})$ of observables $\mathcal{O}({\bf X})$ 
    and a VAR forecaster (\ref{var}). 
    a)-f) : $\mathcal{O}$ vs $\hat{\mathcal{O}}$ from over 20 targets, 
    including d) dislocation junctions and total length of e) screw or  f) $\langle112\rangle/6$ dislocations. 
    Non-scalar a) $\sigma_{\rm xy}$, estimated with $\bar{\bf D}\oplus\bar{\bf V}$\cite{SM}. See also
    $g(r)$ in Fig. \ref{fig:np}.}
    \label{fig:lml}
\end{figure}

\begin{figure}
    \centering
    \includegraphics[width=\columnwidth]{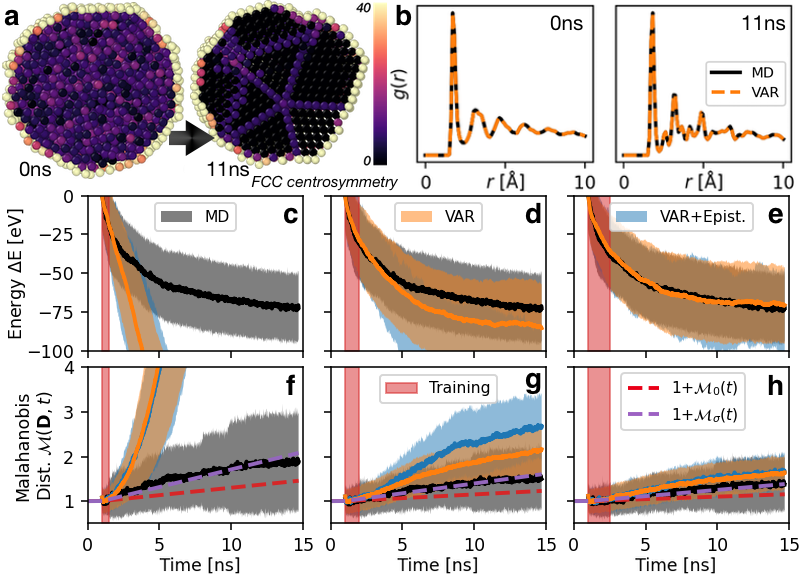}
    \caption{\change{Annealing of Pt nanoparticles. a) Representative structure at 0,11ns. b) The average RDF
    $g(r)$ and the corresponding descriptor estimation\cite{SM}$g(r;\bar{\bf D})$. c)-h) Ensemble data with $M=60$, $\tau_{\rm tr}=0.5,1.0,1.5$ns, training starting at $t=$1ns (left-right, red shade). Mean is solid line, with standard deviation as bands. Black: MD data. Orange: VAR forecasts from 1ns. Blue: VAR forecast with epistemic errors. c)-e) Potential energy change from 1ns mean. f)-h) Mahalanobis distances, MD: $\mathcal{M}({\bf D}_t)$, eq. (\ref{nmd}), forecasts: $\mathcal{M}(t)$, eq. (\ref{tdnmd}). The theoretical lower bounds $\mathcal{M}_0$ (red dash) and $\mathcal{M}_\sigma$ (purple dash) are also shown.}}
    \label{fig:np}
\end{figure}

\textit{Descriptor coarse-graining of microstructure} 
Descriptors\cite{deringer2019machine,goryaeva2020dist_score,Bartok_machine_2017,batatia2022mace}
map atomic coordinates ${\bf X}\in\mathbb{R}^{N\times3}$
to ${\bf D}({\bf X})\in\mathbb{R}^{N\times D}$, where each element 
$[{\bf D}({\bf X})]_{ij}$ takes the local atomic environment of an atom $i$ as input and returns a 
permutation-invariant scalar, vector or tensor depending on the regression target (e.g. energies, forces)\cite{chmiela2017machine,batzner20223}).
Descriptors which approximate a many-body atomic
basis\cite{Thompson_snap_2015,Shapeev_MTP,goryaeva2021,allen2021atomic,lysogorskiy2021performant}
have found use in \textit{linear} estimators
$\hat{\mathcal{O}} \simeq {\bf\Theta}_\mathcal{O}\cdot{\bf D}+\Theta^0_\mathcal{O}$
of some target observable $\mathcal{O}({\bf X})$,
where ${\bf\Theta}_\mathcal{O}\in\mathbb{R}^{D}$ and $\Theta^0_\mathcal{O}\in\mathbb{R}$ are
parameters. For $\mathcal{O}=E$, the atomic potential energy, these can reach state of the art accuracy\cite{batzner20223,bartok2010,Bartok_machine_2017}, often with lower computational cost and simpler 
fitting\cite{goryaeva2021,Thompson_snap_2015,allen2021atomic,lysogorskiy2021performant,Shapeev_MTP}.
The first main result of this contribution is that linear estimators can capture essentially any
microstructural property which could be of relevance to a coarse grained model.
\change{The widely used\cite{LAMMPS} bSO(4) descriptors\cite{bartok2010,Thompson_snap_2015,goryaeva2021,LAMMPS} are used, detailed in the supplementary material (SM)\cite{SM}, summing over all atoms to give the global descriptors}
\begin{equation}
    \bD
    =
    \sum_i{\bf D}_i
    \in\mathbb{R}^D
    ,
    \bar{\bf V}
    =
    \sum_i{\bf X}_i\otimes{\bm\nabla}_{\bf X}{\bf D}_i
    \in\mathbb{R}^{D\times3\times3},
    \label{eq:mm}
\end{equation}
where $\otimes$ is the outer (dyadic) product\footnote{
In practice, the tensor of symmetric matricies $\bar{\bf V}_s\equiv\bar{\bf V}+\bar{\bf V}^\top$
is calculated in current implementations\cite{LAMMPS,goryaeva2021}}. 
Figure (\ref{fig:lml}) shows linear estimators
\begin{equation}
    \hat{\mathcal{O}}(\D) = {\bf\Theta}_{\mathcal{O}}\cdot\D+\Theta^0_\mathcal{O},
    \label{eq:lml}
\end{equation}
applied to dislocation networks in aluminum\cite{SM}, accurately 
capturing a broad range of properties including dislocation junction densities, 
character-dependent line densities and crystal structure content. 
Similar results were found for the nanoparticle ensemble and
a range of dislocated solids in fcc and bcc materials.
Dislocation properties were extracted with \texttt{OVITO-DXA}\cite{stukowski2012automated} 
which has some intrinsic noise due to the discretization parameters. 
\change{It is also possible to capture the radial distribution function (RDF) $g(r)$
by estimating coefficients $\hat{a}_l({\bf D})$ of a basis expansion $g(r)\equiv\sum_la_lu_l(r)$, as shown in Fig. \ref{fig:np}. As found in previous work targeting vibrational entropies\cite{lapointe2020machine,lapointe2022}, 
all predictions were stable under widely varying test/train ratios 
and truncation of training data range.
Matrix-valued observables such as the stress $\mathcal{O}({\bf X})={\bm\sigma}\in\mathbb{R}^{3\times3}$ can be
estimated by building equivariant estimators with $\bar{\bf V}$; 
the simplest ($l=0$\cite{batzner20223}) example is simply $\hat{\mathcal{O}}(\D)=
{\bf\Theta}_{\mathcal{O}}\cdot\bar{\bf V}\in\mathbb{R}^{3\times3}$.}
Examples for the non-scalar shear stress $\sigma_{xy}$ are shown in
figure (\ref{fig:lml}) and the SM\cite{SM}.
However, in the following only $\bar{\bf D}$ is used for forecasting, 
targeting the scalar pressure ${\rm Tr}({\bm\sigma})$, as model parameters are 
scalars and $\bar{\bf D}$ has a metric distance\cite{Thompson_snap_2015}.
Whilst (\ref{eq:lml}) is trained on the global descriptor signal (\ref{eq:mm}),
the same procedure can be applied to spatially-dependent signal from atoms
in some voxel discretization. \change{Further investigation of this spatially dependent signal 
and constraints required for any forecast will be the subject of a future contribution}.\\
The accuracy and scope of (\ref{eq:lml}) has particular relevance for massively parallel workflows,
as only $\bD,\bar{\bf V}$ need to be stored to later extract almost any global observable of interest 
\textit{a posteriori} after training on a small database of stored positions, offering massive 
data compression. \\

\textit{Unimodality and generation of descriptor data} 
As the descriptors have a metric distance, similar microstructures 
will be close in descriptor space. In addition, their distribution in sufficiently 
high dimension can be expected to be unimodal, routinely invoked in active learning schemes \cite{podryabinkin2017active,bernstein2019novo} and more recently in the analysis of defect structures\cite{goryaeva2020dist_score}. 
Evidence for nanoparticle and dislocation ensembles is provided in the SM\cite{SM}. 
It is then simple to \textit{generate} plausible descriptor vectors by 
fitting and sampling a multivariate normal distribution
$\mathcal{N}({\bm\mu},{\bm\Sigma})$ 
to the descriptor dataset. An example of this is shown below in figure \ref{fig:Al}, 
where the observed descriptor initial conditions are densely interpolated, allowing 
the evolution of observable distributions to be monitored.\\

\textit{Resampling and forecasting of descriptor trajectories}
At regular intervals $t_n=n\delta\tau$, $\delta\tau\simeq10$ps, a `snapshot'
is taken by time averaging $\bar{\bf X}_n=\tau^{-1}_D\int_{0}^{\bar{\tau}}{\bf X}(t_n+t){\rm d}t$ 
over a period $\bar{\tau}\simeq20-50$fs to reduce noise from thermal fluctuations\cite{swinburne2014}, 
then calculating descriptor vectors $\D_n=\D(\bar{\bf X}_n)$. A small database of positions 
$\bar{\bf X}_n$ is built by recording $1-5\%$ of snapshots, though positions could be selected
adaptively to maximise training diversity.
An ensemble of $M$ simulations thus produces $M$ discrete time 
trajectories $\{\D_n\}$, which are used to train a 
$P$-state vector autoregressive VAR(P) model\cite{lutkepohl2005new,karlsson2013forecasting}
\begin{equation}
	\D_{n+1} =
	\sum_{p=0}^{p=P-1}
	{\bf T}_{p}\D_{n-p}
	+ {\bf c}
	+ {\bf w}_{n},
	\,
	\langle{\bf w}_n^\top{\bf w}_{m}\rangle = {\bf S}\delta_{nm}.
	\label{var}
\end{equation}
For $P>1$ a Wold transformation\cite{wold1948prediction}
${\bf Z}_{n}=1\oplus\D_{n}\dots\oplus\D_{n-p}\in\mathbb{R}^{1+P\tilde{D}}$
casts (\ref{var}) as a Markovian Ornstien-Uhlenbeck equation\cite{coffey2012langevin} 
${\bf Z}_{n+1} = {\bf T}{\bf Z}_{n}+{\bf W}_n$. The maximum likelihood estimator of 
${\bf T}$ is simply the least squares solution, with ${\bf S}$
determined from the residual covariance\cite{karlsson2013forecasting}. 
To minimize generalization error 
a bagging\cite{breiman1996bagging,lakshminarayanan2017simple,petropoulos2018exploring} 
approach was developed, applying 
Bayesian ridge regression\cite{sklearn} to random overlapping subsets. 
Results were stable under  10-40 subsets each with 10-40\% coverage, giving
epistemic uncertainties $\delta{\bf T},{\delta{\bf S}}$ from the covariance across subsets. Training is robust and requires only a few CPU minutes, a key advantage over (RNN/LSTM) neural networks\cite{yu2019,vlachas2021accelerated} 
or neural differential equations\cite{kidger2020neural} which require significant resources, regularisation/correction schemes\cite{fu2022simulate},
and limited in practice to data dimension $\tilde{D}<10$\cite{liu2019ensemble,kidger2020neural}. 
A Chapman-Komologorov test\cite{mardt2018vampnets} for the transfer matrix ${\bf T}$ is provided in the SM\cite{SM}, but in practice the light computational demand also permits a 
convergence test of model architecture by increasing $P$\cite{SM}.\\
\textit{Deriving a forecast uncertainty}
Practical application of (\ref{var}) requires a robust measure of forecast uncertainty\cite{perez2015,swinburne2020automated,garmon2022resource,schaarschmidt2022workflow,zhu2021fully,andrews2022forecasting}, which should be larger for configurations further from the training data independent of epistemic errors. 
This is particularly relevant to the non-stationary dynamics of material deformation. As uncertainty to previously unseen macroscopic changes is clearly not quantifiable\cite{swinburne2020automated},the following bound is conditional on the simulation \textit{ensemble} remaining unimodal and not undergoing macroscopic changes.
Many extrapolation grade estimators have been developed for active learning of energy models\cite{bartok2010,podryabinkin2017active,goryaeva2020dist_score,dadapy,zeni2022exploring};
here, the Mahalanobis outlier distance\cite{mahalanobis1936generalized} is 
used for the unimodal descriptor distribution\cite{SM,goryaeva2020dist_score}.
With training data mean ${\bm\mu}_{\rm tr}$ and covariance ${\bm\Sigma}_{\rm tr}$ estimated
via a shinkage estimator\cite{chen2010shrinkage}, the squared Mahalanobis distance reads
\begin{equation}
    \mathcal{M}(\D) =
    \left[\D-{\bm\mu}_{\rm tr}\right]
    {\bm\Sigma}_{\rm tr}^{-1}
    \left[{\bf D}-{\bm\mu}_{\rm tr}\right]/{\tilde{D}}
    \label{nmd}.
\end{equation}
Importantly, (\ref{nmd}) is independent of the VAR(P) forecast model (\ref{var}); 
points drawn from a low density region of $\rho_{\rm tr}$ will 
have a large Mahalanobis distance, even if epistemic uncertainties $\delta{\bf T}$ are small. 
At long forecasting times, (\ref{var}) will reach its high dimensional 
steady state\cite{SM}, with $\langle\mathcal{M}\rangle$ constant.
However, model parameters cannot be assumed static, with a time dependence bounded from below by $1/\tau_M=1/(M\tau_{\rm tr})$\cite{swinburne2018b}, 
where $M$ is the ensemble size and $\tau_{\rm tr}$ training duration. 
This drift can be estimated by propagating epistemic uncertainty in the steady state to an uncertainty $\sigma^2_{\mathcal{M}}$ in $\mathcal{M}(\D)$, 
which 
should be accumulated\cite{SM}, leading to an additional linear growth in (\ref{nmd}) of
\begin{equation}
    \mathcal{M}(t_n) = \langle\mathcal{M}(\D_{n})\rangle + \mathcal{M}_\sigma(t_n),\quad
    \mathcal{M}_\sigma(t_n)\geq\mathcal{M}_0(t_n).
    \label{tdnmd}
\end{equation}
where $\mathcal{M}_\sigma(t)=\sigma^2_{\mathcal{M}}t/\delta\tau$
and $\mathcal{M}_0(t)=t/\tau_M$.
Equation (\ref{tdnmd}) is the main theoretical result of this contribution, 
an uncertainty metric for forecasting via (\ref{var}). An approximate parallel 
efficiency is implied by $\eta=\tau_{\rm pred}/(M\tau_{\rm tr})$, 
where $\mathcal{M}(\tau_{\rm pred})\equiv\mathcal{M}_0(M\tau_{\rm tr})=2$,
giving $\eta=1$ when $\mathcal{M}(t_n)=\mathcal{M}_0(t)$.

\textit{Annealing of Pt Nanoparticles}
Metallic nanoparticles are important functional materials for catalysis;
50-150 atom clusters have been extensively studied in simulations\cite{wales_energy_2003,baletto2002crossover,huang2017,huang2018direct}, 
but for large sizes and high temperatures the landscape of energy minimia
is vast and insufficiently metastable for current acceleration methods\cite{huang2018direct}. The current application to $M=$60
4000-atom EAM-Pt\cite{liu1991eam} nanoparticles at 900K is thus
untreatable with existing methods.\\

The initial structure was formed by quenching from the liquid state and annealing for 
100ps to give a highly disordered but predominantly fcc structure ($c_{\rm FCC}\simeq0.5$).
Descriptor trajectories were extracted every 1.5 ps, with a full structural analysis undertaken 
every 100ps, though the dataset was sparsified by taking $\delta{\tau}=15$ps and removing intermediate
snapshots. Autoregressive models (\ref{var}) were constructed
with $P=1-3$ and $\tau_{\rm tr}$=0.5, 1.0 or 1.5ns, with $P=1$ shown.
Generated trajectories were launched from the start of the training stage,
meaning the observed trajectories were resampled and then forecasted.
Figure \ref{fig:np} displays the ensemble simulation data, model predictions
and epistemic errors for the formation energy, \change{the RDF $g(r)$}
and the Mahalanobis uncertainty (\ref{tdnmd}). The RDF reflects the significant 
growth in FCC crystal structure, as can also be directly extracted through 
estimation of $c_{\rm FCC}$, as shown in the SM\cite{SM}.
MD data used $\mathcal{M}({\bf D})$, eq. (\ref{nmd}),
which closely follows the theoretical lower bound $\mathcal{M}_\sigma(t)$.
Whilst the prediction error systematically improves as training data increases
in duration and diversity, of central importance is that this is reflected in 
the magnitude of $\mathcal{M}(\D,t)$, confirming that predictions are reliable. 

\begin{figure}
    \centering
    \includegraphics[width=\columnwidth]{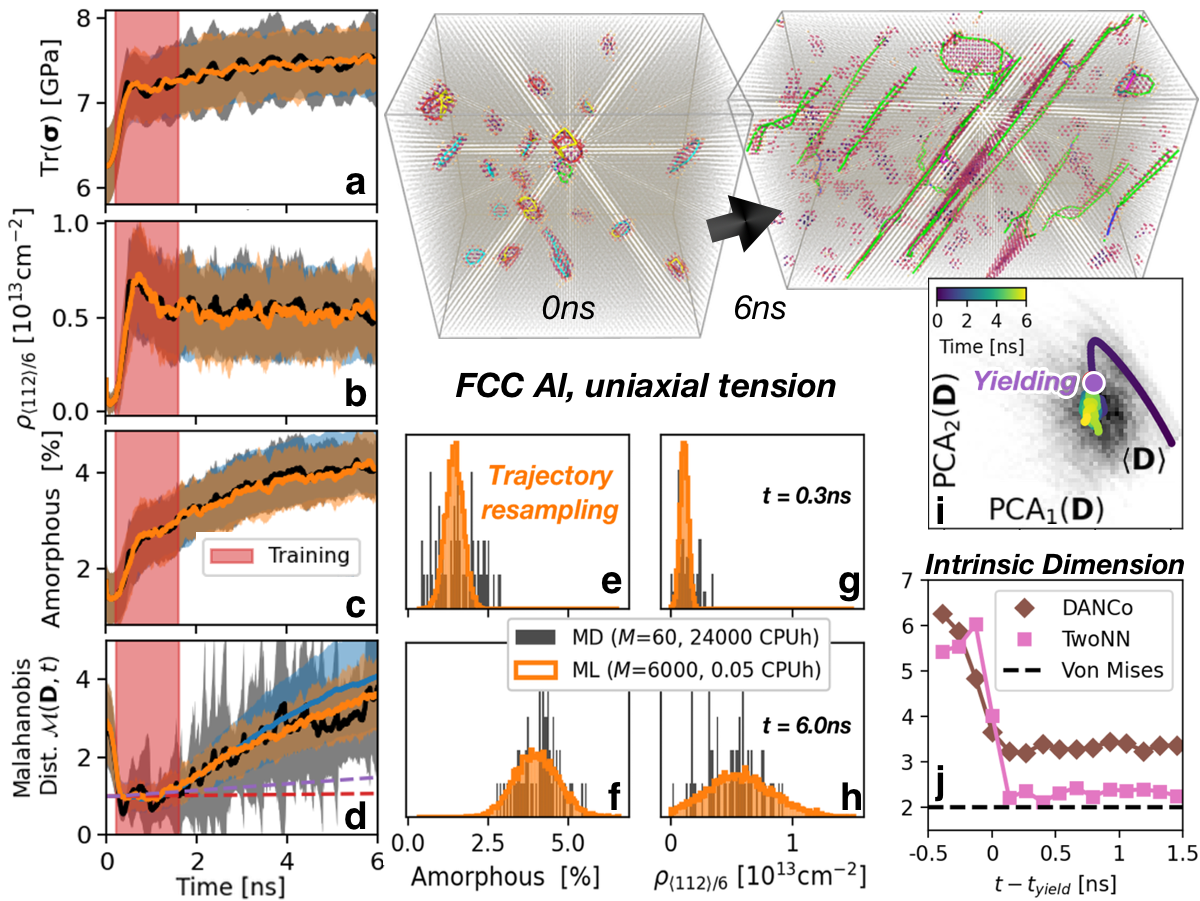}
    \caption{\change{Yielding of dislocation networks under uniaxial tension in Al. 
    Color scheme follows figure \ref{fig:np}. Forecasts are from 0.2ns, with
    $\tau_{\rm tr}=1.2$ns and $P=5$. Observable plots a) Pressure, 
    b) $\langle112\rangle/6$ dislocation density and c) amorphous 
    content. d) Mahalanobis distance. 
    e)-h) Trajectory resampling from 0.3ns (pre-yield), 
    with $100\times$ larger ensemble. i) PCA analysis of the ensemble mean 
    $\langle{\bar{\bf D}}\rangle$, clearly showing a localization on yield.
    Individual trajectories shown as histogram in grayscale.
    j) ID of the descriptor manifold, estimated via the TwoNN\cite{facco2017estimating} 
    and DANCo\cite{scikit-dim} methods. Both show a collapse on yielding, 
    but remain above the Von Mises lower bound.} }
    \label{fig:Al}
\end{figure}

\textit{Yielding of Al under uniaxial tension}
\change{Dislocations the agents of plastic deformation and form dense networks
under irradiation\cite{mason2020observation} or extended 
loading\cite{lavenstein2020heterogeneity}, with atomic simulations 
still producing unexpected atomic mechanisms even in extensively studied pure 
materials\cite{bertin2022enhanced}. Understanding yield
is a central goal of metal physics\cite{baggio2019landau}, 
but atomic analysis is challenging
as collective variables are elusive\cite{baima2022} and the dynamics
are insufficiently metastable or stationary for the methods discussed above. }
An ensemble of $M=60$ dense dislocation networks were formed in 
an EAM model of Al\cite{onat2013optimized} by creating simulation boxes
of around $1.5\times10^5$ atoms, orientated to $[10\bar{1}],[1{1}1],[1\bar{2}1]$,
with populations of interstitial loops. The initial dislocation densities 
spanned the typical MD range\cite{zepeda2017probing,bertin2022enhanced} of
$\rho_{\rm dis}\in[10^{11},10^{13}]{\rm cm}^{-2}$. Uniaxial tension was applied 
at a rate $\dot{\epsilon}_{xx}=1\times10^8{\rm s}^{-1}$ along $[10\bar{1}]$, 
allowing other supercell dimensions to relax\cite{zepeda2017probing}.
\change{The SM\cite{SM} shows application to cyclic shear loading}.
The results are summarized in figure \ref{fig:Al}a)-d), using the
linear estimators (\ref{eq:lml}). 
\change{Increasing $\tau_{\rm tr}$ decreased error and uncertainty- convergence tests 
showed optimal results with $P=5$\cite{SM}. Training only on pre-yield structures led to 
unstable forecasts as yield is characterized by a qualitative change in the 
descriptor manifold as detailed below. However, $\mathcal{M}(\bar{\bf D},t)$ also 
diverged at the yield point, clearly indicating that more training data is 
required. This again demonstrates the utility and critical importance of a 
forecast uncertainty to asses data-driven predictions.}
Resampling allows for ensembles to be increased by orders of magnitude for 
negligible CPU effort, giving smooth microstructure distributions as shown in figure \ref{fig:Al}e)-h).
Initial descriptor states were generated as described above 
from $\rho_{\rm tr}\simeq\mathcal{N}({\bm\mu}_{0},{\bm\Sigma}_{0})$, 
fit from the descriptor ensemble at times 0.3-0.31ns. 
\change{The forecasted ensemble captures multiple 
important microstructural evolutions that, whilst known 
for this well-studied system\cite{zepeda2017probing}, 
confirm the accuracy of the VAR approach. Forecasts
correctly predict the growth of amorphous atomic environments 
due to defect production under continued loading\cite{marian2004dynamic}, 
the expected sharp peak in HCP content at yield, accompanied by a growth,
peak then steady state in the number of dislocation junctions (see SM\cite{SM}). Distributions can tighten or widen, here indicating the evolution in dislocation character- initial populations of $\langle100\rangle/3$ Hirth dislocation loops decay to a tight distribution close to zero upon loading, accompanied by 
an emergence of a broad, stable distribution of $\langle112\rangle/6$ dislocation lines
that carry the plastic flow\cite{zepeda2017probing}. The joint stability of 
junctions, dislocation density and stress is consistent with a Kocks-Mecking steady state\cite{sills2018dislocation}.}
\change{
The descriptor data is highly sensitive to yielding; figure \ref{fig:Al}i)
shows the first two PCA components of the ensemble mean $\langle D\rangle$,
revealing a clear localization after yielding which can be easily classified (see SM\cite{SM}).
Theoretical models for yielding invoke the concept of a yield surface in 
5D stress space\cite{hirth}, which for metallic systems is typically the Von Mises 
yield surface, isosurfaces of the $J_2$ invariant with intrinsic dimension
(ID)\cite{scikit-dim,dadapy} of 2. Yielding is thus expected to be accompanied by an 
abrupt drop in the intrinsic dimensionality of the stress trajectory. 
As stresses are essentially deterministic from descriptors, the descriptor ID is an upper bound to
the yield surface ID.
Two empirical ID estimators\cite{scikit-dim,facco2017estimating,bac2021scikit} 
were applied to the full descriptor data $\bar{\bf D}\oplus\bar{\bf V}$. 
Figure \ref{fig:Al}j) shows the estimated ID collapses from around 5-7 to 
around 2-3 on yield. Although these typically underestimate\cite{scikit-dim}, this
is consistent with a Von Mises ID lower bound of 2. Furthermore, 
this indicates the existence of a yield manifold in descriptor space,
generalizing the yield surface concept to a much richer description of 
microstructure than stress alone. It is speculated that the yield manifold provides a route for data-driven construction of advanced structure-property relationships.}\\
\textit{Conclusions}
\change{
This contribution has promoted scalar and matrix-valued descriptors as a compressed 
representation of atomic microstructure ideal for analysis, resampling and forecasting of simulations,
with a robust forecast uncertainty derived using outlier distances. Analysis of the descriptor manifold 
indicates the existence of a generalized yield surface, a promising direction for future research, 
alongside the use of forecasting in autonomous resource allocation\cite{perez2015,swinburne2020automated,garmon2022resource,schaarschmidt2022workflow,zhu2021fully,andrews2022forecasting} 
and extension to a spatially dependent, fully equivariant descriptor framework.}
\section{Acknowledgments}
I thank M-C Marinica and L Truskinovsky for stimulating discussions, an anonymous referee
for careful reading of the manuscript, the ANR grant ANR-19-CE46-0006-1, IDRIS allocations A0090910965, A0120913455, and 
Euratom grant No 633053.

\newpage
\onecolumngrid
\newpage

\section{Supplementary material for `Coarse-graining and forecasting atomic simulations of solids with descriptor functions`}

\subsection{Choice of descriptor function basis}
Descriptors which approximate a many-body atomic
basis\cite{Thompson_snap_2015,Shapeev_MTP,goryaeva2021,allen2021atomic,lysogorskiy2021performant}
have found use in \textit{linear} (LML) models
$\mathcal{O} = {\bf\Theta}_\mathcal{O}\cdot\sum_i{\bf D}_i$ of some target $\mathcal{O}$,
where ${\bf\Theta}_\mathcal{O}\in\mathbb{R}^D$ is a vector of parameters
\footnote{offsets are absorbed into the first entry of the per-atom descriptor ${\bf D}_i$ (i.e. ${\rm D}_{i1}=1$)}.
A number of possible basis functions have been proposed; an approximate classification
can be made into `compact` spectral expansions of 50-200 terms\cite{goryaeva2021,Thompson_snap_2015}
and `non-compact` polynomial expansions of 1000-10000 terms, which offer greater accuracy
but must be carefully regularised to avoid overfitting\cite{allen2021atomic,lysogorskiy2021performant,Shapeev_MTP}.
In this work, the bSO(4) bispectral descriptor functions are used\cite{bartok2010,Thompson_snap_2015,goryaeva2021},
as implemented in the \texttt{MILADY} potential package, first 
introduced as part of the \texttt{SNAP} family of LML potentials~\cite{Thompson_snap_2015}.
Briefly, let ${\rm B}_{ji}({\bf X}),j\in[0,{\rm N_B}]$ be the ${\rm N_B}$
bispectral components for an atom $i$, along with a constant 
component ${\rm B}_{0i}\equiv1$. Only neighboring atoms 
within the cutoff distance (here 4.7 ${\text \AA}$) are included in the 
descriptor function calculation, which for the atom $i$ reads
\begin{equation}
    {\bf D}_i({\bf X})
    =
    \sum_i
    {\bigoplus}_{j} {\rm B}_{ji}({\bf X}) \in\mathbb{R}^{{\rm N_B}}
    ,
    \label{eq:desc}
\end{equation}
where $\oplus$ indicates concatenation, giving $({\rm N_B}+1)$ components.
The number of bispectrum components ${\rm N_B}$ is determined by an angular moment parameter 
$j_{max}$, here set to $4$, giving ${\rm N_B}=55$.

\change{
\subsection{Descriptor estimation of correlation functions}
As discussed in the main text, it is also possible to predict the value of correlation functions such 
as the radial distribution function $g(r)$, which could be written as $g(r;{\bf X})$, as it is
deterministic from a given set of positions ${\bf X}$. Our goal is to make a basis expansion
\begin{equation}
    g(r;{\bf X}) = \sum_l a_l({\bf X}) u_l(r), 
\end{equation}
where $u_l(r)$ are linearly independent. As the coefficients $a_l({\bf X})$ are deterministic
scalar functions of $\bf X$ they are valid targets for estimation by a descriptor estimator $\hat{a}_l({\bf D})$.
In practice, $g(r;{\bf X})$ is evaluated up to a multiplicative constant as a histogram with a vector of $H$ counts ${\bf g}({\bf X})\in\mathbb{Z}^H_+$ in $H$ bins with centers ${\bf r}$. 
Fixing ${\bf r}$, an appropriate discrete basis ${\bf u}_l\in\mathbb{R}^H$ can be found through singular value decomposition
of the rectangular matrix of $T$ training points 
${\bf G}=[{\bf g}({\bf X}_1),{\bf g}({\bf X}_2),\dots]\in\mathbb{R}^{T\times H}$. 
It is then simple to learn the coefficient estimators $\hat{a}_l({\bf D})$
for as many singular vectors as required, which in practice was found to be 4 or 5. It is noted that $g(r)$ was
only considered as an observable after simulations has been run; this is a key advantage of the current approach, 
as adding $g(r)$ did not require any new data generation, only 
estimator training on the 1\% of retained positions then prediction on the full descriptor dataset. This ability
to select coarse-graining targets \textit{a posteriori} is a central advantage of the current method.
}

\newpage
\subsection{Regression of matrix-valued observables}
\begin{figure}[!h]
    \centering
    \includegraphics[width=0.49\columnwidth]{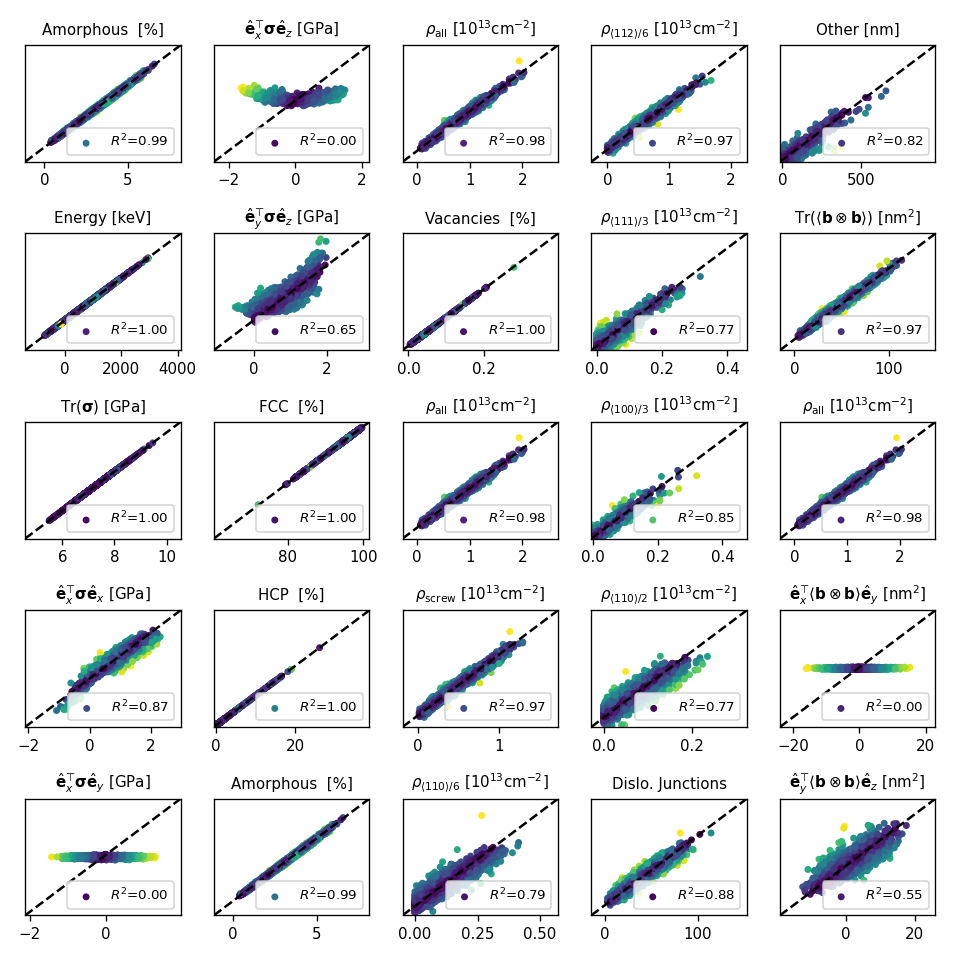}
    \includegraphics[width=0.49\columnwidth]{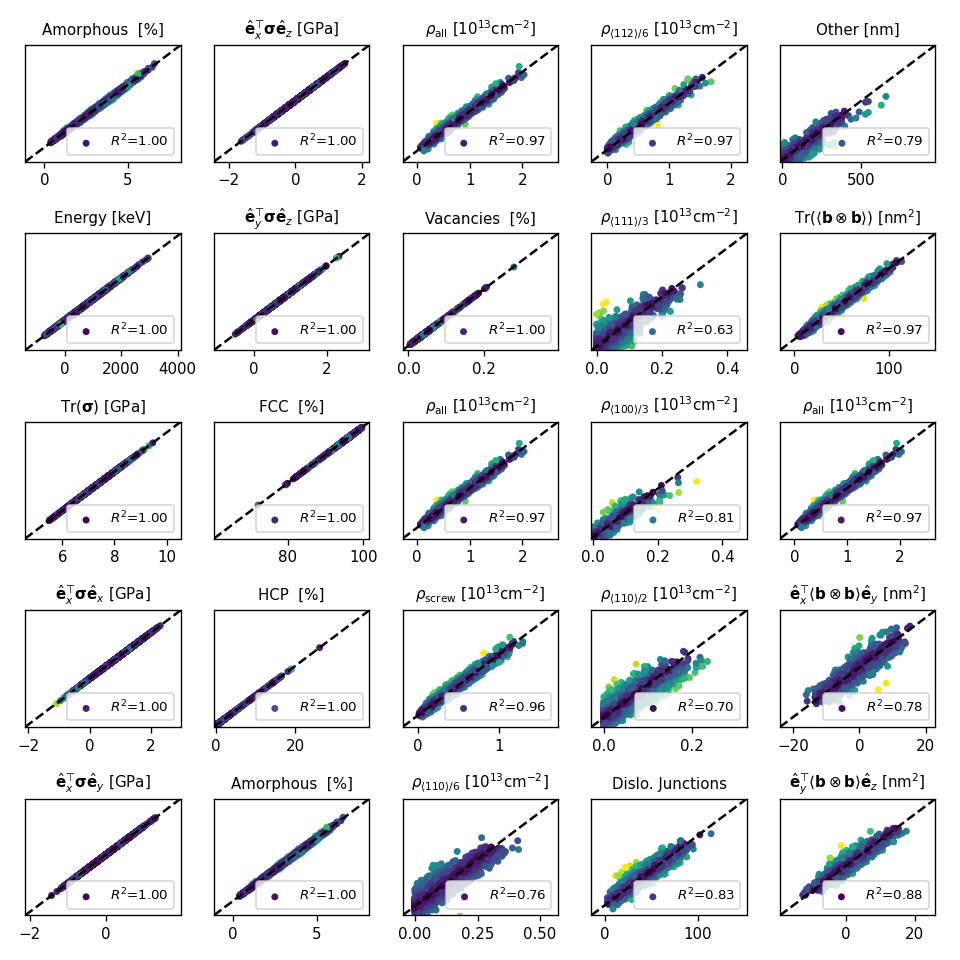}
    \caption{Linear models with 50 components from PCA of $\bar{D}$ (left) and 150 components from PCA on 
    $\bar{\bf D}\oplus\bar{\bf V}$. Data is for the Al under uniaxial loading presented in the main text, 
    with the same training regime.}
    \label{fig:all_lml}
\end{figure}

\begin{figure}[!h]
    \centering
    \includegraphics[width=0.4\columnwidth]{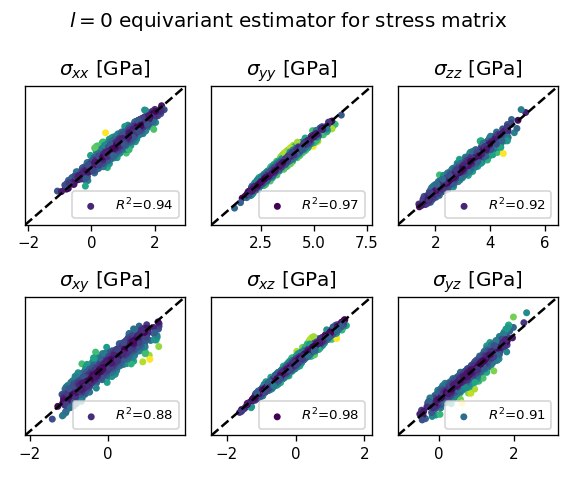}
    \caption{Matrix-equivariant ($l=0$ coefficients) model for the stress $\hat{\bf\sigma}={\bf\Theta}_\sigma^\top\bar{\bf V}$. Data is for the Al under uniaxial loading presented in the main text.}
    \label{fig:eq_stress}
\end{figure}

As discussed in the main text, the descriptor vectors are then summed to give the extensive vectors
$\bar{\bf D}=\sum_i{\bf D}_i$, $\bar{\bf V}=\left[\sum_i{\bf X}_i\otimes{\bm\nabla}_{\bf X}{\bf D}_i\right]$
where $\hat{\bf e}_\alpha,\hat{\bf e}_\beta\in\mathbb{R}^3$ are Cartesian axes and $\otimes$
is the outer (dyadic) product. In principle, the scalar $\bar{\bf D}$ are sufficient to regress scalar projections 
of matrix quantities such as shear stresses $\sigma_{xy}={\bf e}_x\cdot{\bm\sigma}\cdot{\bf e}_y$.
However, whilst not used in the main text, it is possible to form scalar projections 
$\bar{\bf V}_{\alpha\beta} = \bar{\bf V}:{\bf e}_\alpha\otimes{\bf e}_\beta$. It is then possible 
to regress shear stress components $\sigma_{xy}={\bf e}_x\cdot{\bm\sigma}\cdot{\bf e}_y$.
The effect of this can be seen in figure \ref{fig:all_lml}.
It is possible to form matrix-equivariant linear models using only scalar coefficients 
$\hat{\mathcal{O}}={\bf\Theta}^\top\bar{\bf V}$, shown in figure \ref{fig:eq_stress}.
Use of $\bar{\bf D}\oplus\bar{\bf V}$ as in figure \ref{fig:all_lml} gave slightly superior results; 
future work will investigate equivariant use of $\bar{\bf D}$ e.g. 
$\hat{\mathcal{O}}={\bf\Theta}^\top\bar{\bf V} + \bar{\bf D}^\top{\bm\Phi}\bar{\bf V}$.

\newpage

\begin{figure}[!h]
    \centering
    \includegraphics[width=0.7\columnwidth]{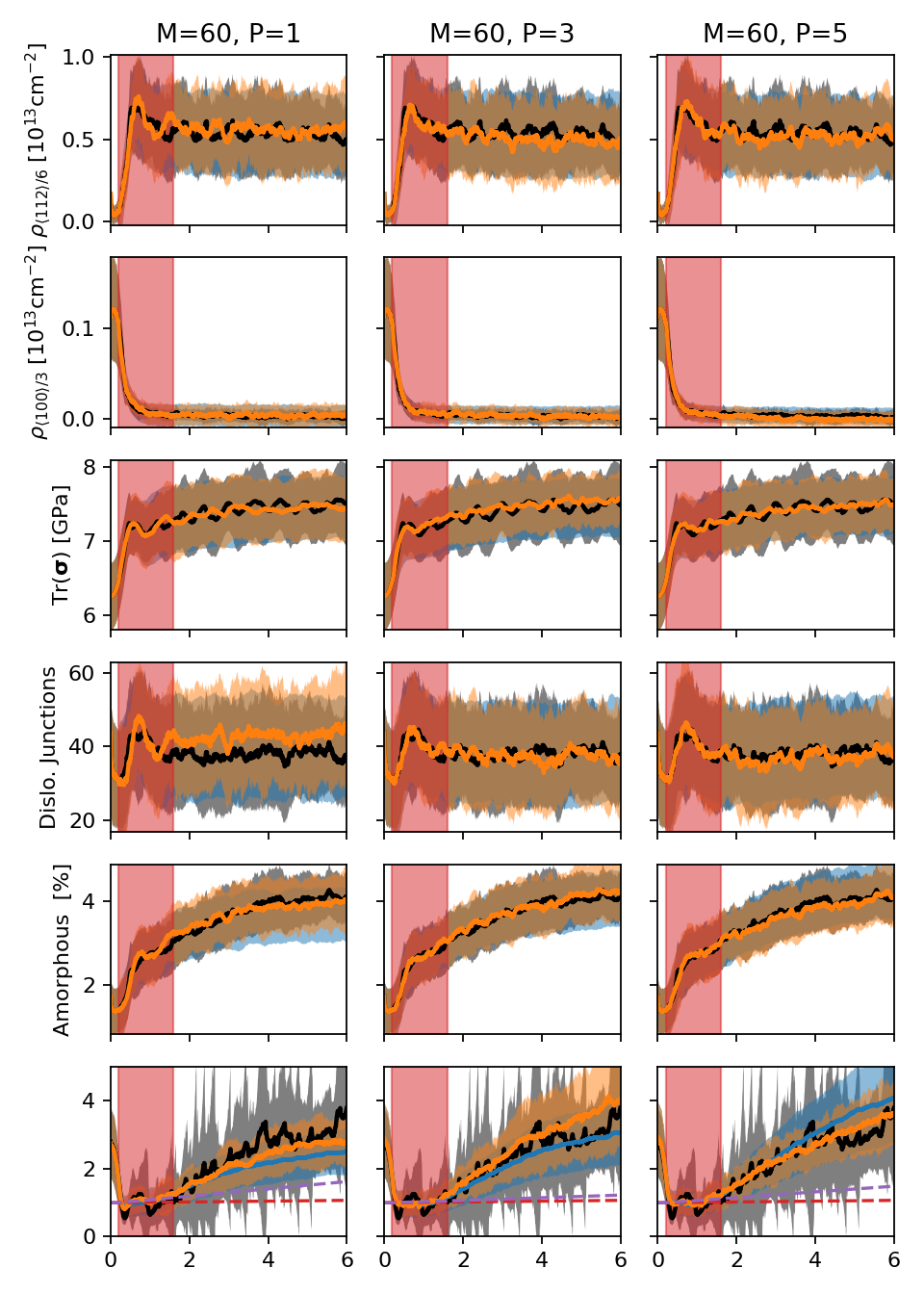}
    \caption{Convergence test of VAR approach for Dislocation networks in Al under uniaxial tension, as presented in the main text. 
    Left-right: VAR models with $P=1,3,5$. $P=5$ was used in the main text. }
    \label{fig:Al_P}
\end{figure}

\change{\subsection{$P$-convergence of VAR models}
In addition to Chapman-Komologorov tests (see below) it is also simple to study convergence in model predictions with the memory $P$. 
As can be seen in figure \ref{fig:Al_P} , the predictions are stable and mildly improve with increasing $P$.}

\subsection{Chapman-Komologorov tests}
VAR models with $P=1$ are equivalent to extended dynamic mode decomposition\cite{williams2015data}, 
for which a Chapman-Komologorov test is possible. In general, in implied timescales $\omega = \ln|\lambda_\tau|/\tau$ should be a constant. As can be seen, this is satisfied for $\tau>10$ps.
\begin{figure}[!h]
    \centering
    \includegraphics[width=0.45\columnwidth]{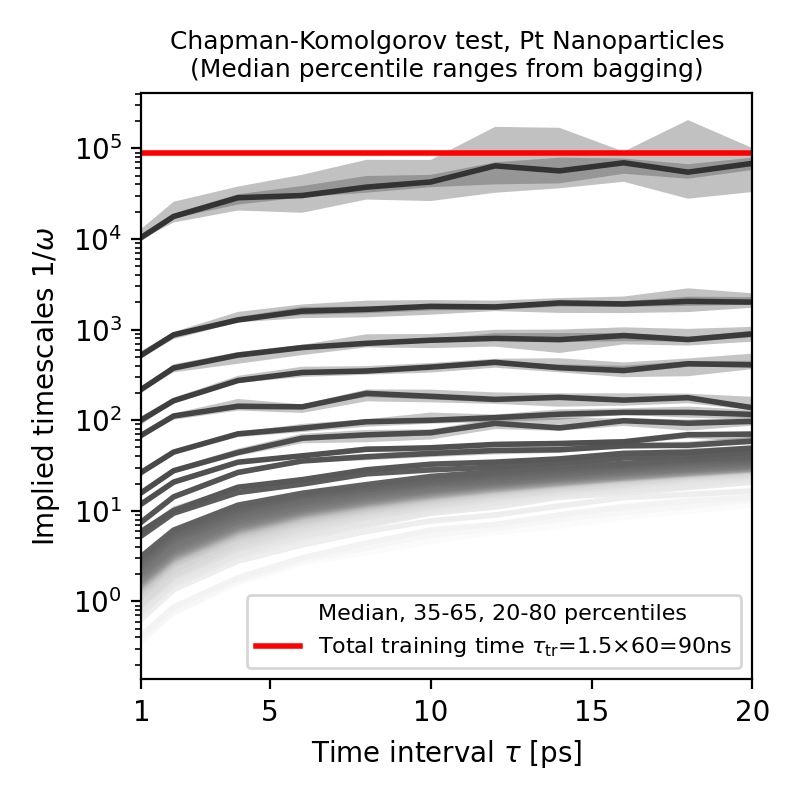}
    \includegraphics[width=0.45\columnwidth]{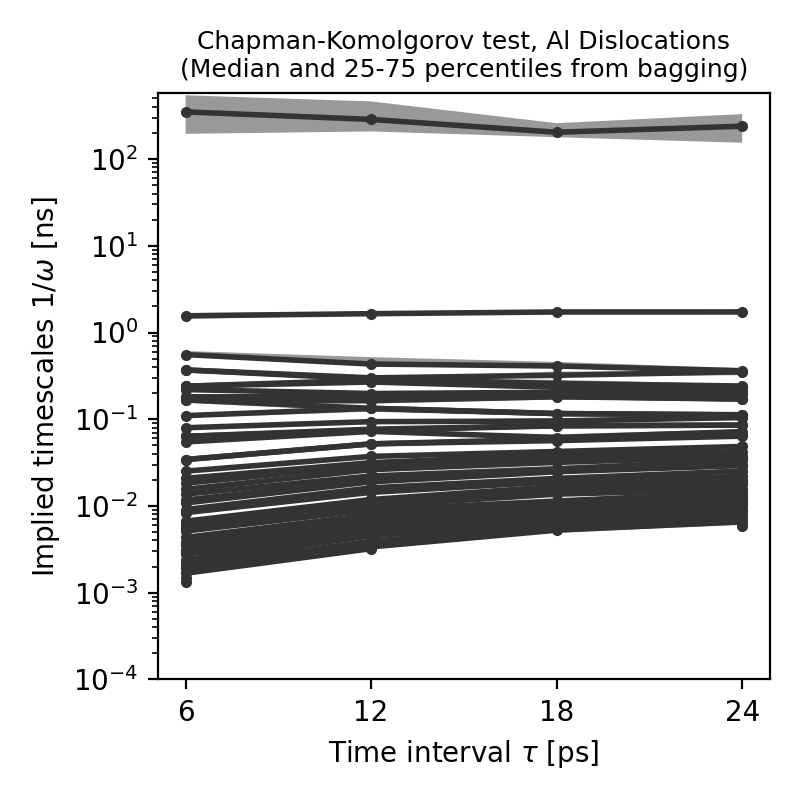}
    \caption{Chapman-Komologorov test, $\omega = \ln|\lambda_\tau|/\tau$, for the Pt nanoparticle and Al dislocation ensembles. The implied 
    timescales $1/\omega$ are constant for $\tau\geq$10ps, which is the employed interval.}
    \label{chap}
\end{figure}

\subsection{Steady state of autoregressive models}
The VAR(P) vector autoregressive model reads\cite{lutkepohl2005new}
\begin{equation}
	\tilde{\bf D}_{n} =
	\sum_{p=1}^P
	{\bf T}_{p}\tilde{\bf D}_{n-p}
	+ {\bf c}
	+ \delta{\bf W}_{n},
	\quad
	\langle \delta{\bf W}_n^\top\delta{\bf W}_{m}\rangle = {\bf S}\delta_{nm}.
	\label{var}
\end{equation}
Consider the vector ${\bf Z}_{n-1} = [\tilde{\bf D}_{n-1},\dots,\tilde{\bf D}_{n-p}]$.
It satisfies the extended VAR(1) model via a Wold transformation\cite{wold1948prediction}
\begin{equation}
	{\bf Z}_{n} =
	\left[
	\begin{matrix}
		{\bf T}_{1} & \dots & {\bf T}_{p} \\
		\mathbb{I} & {\bf 0} & \dots \\
		{\bf 0} & \mathbb{I} & \dots \\
		{\bf 0} & \vdots & \dots \\
		\dots &  \mathbb{I} & {\bf0}
	\end{matrix}
	\right]
	{\bf Z}_{n-1}
	+
	\left[
	\begin{matrix}
		{\bf c} + \delta{\bf W}_{n} \\
		{\bf 0} \\
		{\bf 0} \\
		\vdots \\
		{\bf 0} \\
	\end{matrix}
	\right]
	=
	{\bf T_Z}
	{\bf Z}_{n}
	+
	\delta{\bf Y}_{n}.
\end{equation}
We can now solve for the steady state as for any other VAR(1), or
Ornstien-Uhlenbeck equation\cite{coffey2012langevin},
with a Gaussian steady state
\begin{equation}
	\lim_{t\to\infty} \rho({\bf Z},t) = \mathcal{N}({\bf Z}_\infty,{\bf Q}_\infty)
\end{equation}
where
\begin{equation}
	\left[\mathbb{I}-{\bf T_Z}\right]{\bf Z}_\infty\equiv{\bf c}
	,\quad
	{\bf S}={\bf L}^\top{\bf L}
	,\,
	\left[\mathbb{I}-{\bf T_Z}\right]{\bf V} = {\bf L}
	,\quad
	{\bf Q}_\infty = {\bf V}^\top{\bf V},
\end{equation}
where we express the mean and variance in a form which allows for a
null space of $\left[\mathbb{I}-{\bf T_Z}\right]$ in the case of inadequate
training data or underregularisation\cite{Rasmussen}.
\newpage

\subsection{Unimodality of descriptor distributions}

\begin{figure}[!h]
    \centering
    \includegraphics[width=\columnwidth]{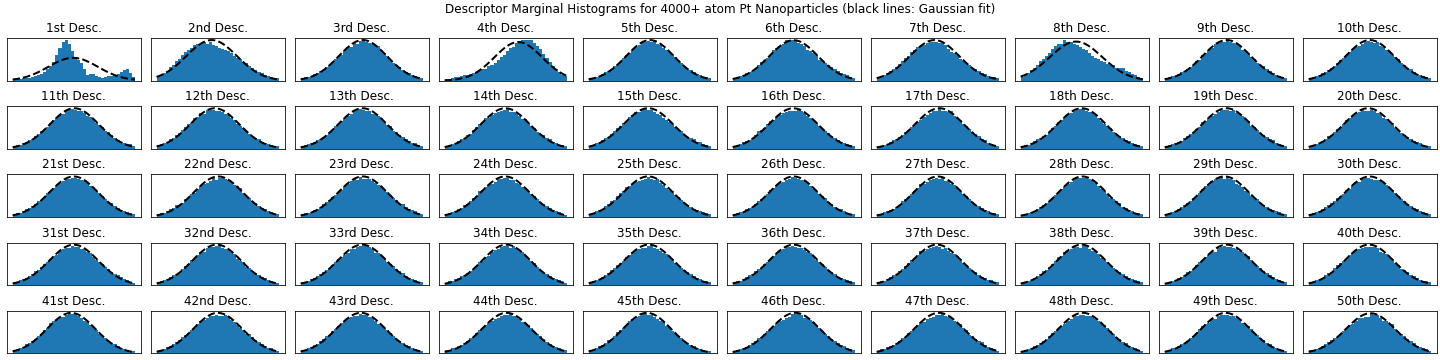}
    \includegraphics[width=\columnwidth]{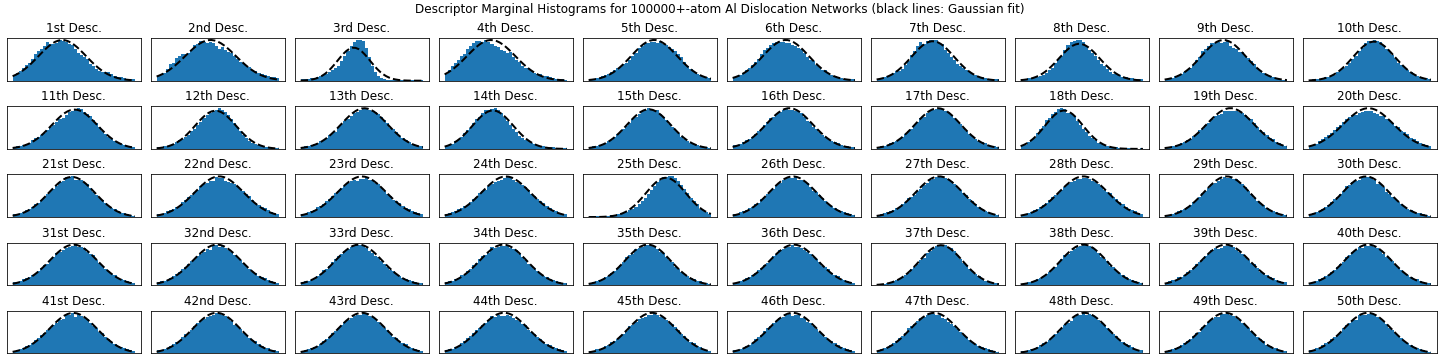}
    \caption{Descriptor histograms for the Pt nanoparticles (top) and Al histograms (bottom).}
    \label{fig:unimod}
\end{figure}

As discussed in the text, the error metrics we employ rely on approximately 
unimodal descriptor distrinbutions. Figure (\ref{fig:unimod}) shows the descriptor datasets used for 
the forecasting model, the raw descriptor data reduced to 50 dimensions and whitened (decorrelated)
through application of the PCA technique\cite{Rasmussen}. As can be seen, the histograms are unimodal
to a high degree of approximation. 
The only slight deviation came from one component for the Pt nanoparticles, but we attribute 
this primarily to the relatively small sample size of 60 simulation workers in a small atomic system. 
For the Al dislocation system, the unimodality is clear. 
\subsection{Uncertainty propagation to the Mahalanobis distance}
The bagging regression proceedure detailed in the main text returns expectation values
and covariances for model parameters ${\bf T_Z},{\bf c},{\bf S}$.
In practice, we decompose ${\bf S}={\bf X}^\top{\rm Diag})({\bm\lambda}){\bf X}$
and only consider covariance of the positive eigenvalues ${\bm\lambda}$. 
Epistemic uncertainties in ${\bf T_Z}$ where found to have negligable influence and thus 
are not considered here. 
These parameter uncertainties, expressed by the covariances ${\bm\Sigma}_{\bf c}$ and 
${\bm\Sigma}_{\bm\lambda}$ of ${\bf c}$ and ${\bm\lambda}$, will induce a random perturbation
on the square Mahalanobis distance 
\begin{equation}
    \mathcal{M}(\D) =
    \left[\D-{\bm\mu}_{\rm tr}\right]
    {\bm\Sigma}_{\rm tr}^{-1}
    \left[{\bf D}-{\bm\mu}_{\rm tr}\right]/{\tilde{D}}
    \label{nmd}.
\end{equation}
As discussed in the main text, as this uncertainty always increases $\mathcal{M}$ it cannot be assumed to be a 
mean zero fluctuation and thus should be accumulated, 
leading to a linear growth in time. It is simple to propagate this uncertainty to 
$\mathcal{M}(\D)$:
\begin{equation}
    \sigma^2_{\mathcal{M}} = 
    {\rm Tr}\left(\left[\mathbb{I}-{\bf T_Z}\right]^{-1}{\bm\Sigma}_{\bf c}{\bm\Sigma}_{\rm tr}^{-1}\right)
    /{\tilde{D}}
    +
    {\rm Tr}\left(\left[\mathbb{I}-{\bf T_Z}\right]^{-1}{\bf X}^\top{\bm\Sigma}_{\bm\lambda}{\bf X}{\bm\Sigma}_{\rm tr}^{-1}\right)
    /{\tilde{D}}
\end{equation}
In practice, we ensure that $\sigma^2_{\mathcal{M}}>\delta\tau/\tau_M$ though the mapping 
$\sigma^2_{\mathcal{M}}\to1/(1/\sigma^2_{\mathcal{M}}+\tau_M/\delta\tau)$, though as can be 
seen in the figures the addition of this has only a very minor influence on the expected uncertainty.

\change{\subsection{Yielding of dislocation networks under cyclic shear}
An ensemble of $M=60$ dense dislocation networks were formed in 
an EAM model of Al\cite{onat2013optimized} by creating simulation boxes
of around $1.5\times10^5$ atoms, orientated to $[10\bar{1}],[1{1}1],[1\bar{2}1]$,
with populations of interstitial loops. The initial dislocation densities 
spanned the range typically found in MD\cite{zepeda2017probing,bertin2022enhanced}, 
$\rho_{\rm dis}\in[10^{11},10^{13}]{\rm cm}^{-2}$. 
Ensembles were subjected to cyclic shearing at a rate 
$|\dot{\epsilon}_{xy}|=1\times10^8{\rm s}^{-1}$ up to $|\epsilon_{xy}|=0.1$
with a sawtooth profile for a total of two cycles. As can be seen, the initially 
disperse dislocation densities across the ensemble tightens after a few cycles, 
as the driven steady state emerges.}
\begin{figure}[!h]
    \centering
    \includegraphics[width=0.9\columnwidth]{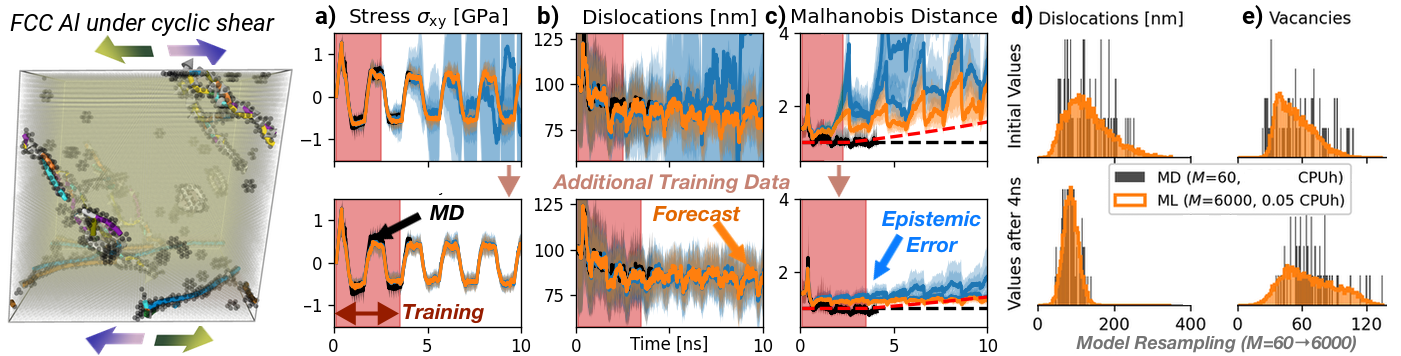}
    \caption{Cyclic shearing of dislocation networks in Al. Left: Representative simulation supercell. a)-c): Forecasts using 2.5ns (top) and 3ns (bottom) for training and resampling, showing a) shear stress, b) total dislocation density and c) outlier distance. For lower training data the outlier distance is clearly seen to spike at yeild points, these being the least sampled configurations. d)-e): resampling, increasing the effective ensemble size one hundredfold. The total dislocation density distribution d) noticably tightens. demonstrating a convergence to a driven steady state, whilst the vacancy distribution e) slightly widens, most likely due to defect generation at high strain rates.}
    \label{fig:Al_cc}
\end{figure}

\begin{figure}[!h]
    \centering
    \includegraphics[width=0.5\columnwidth]{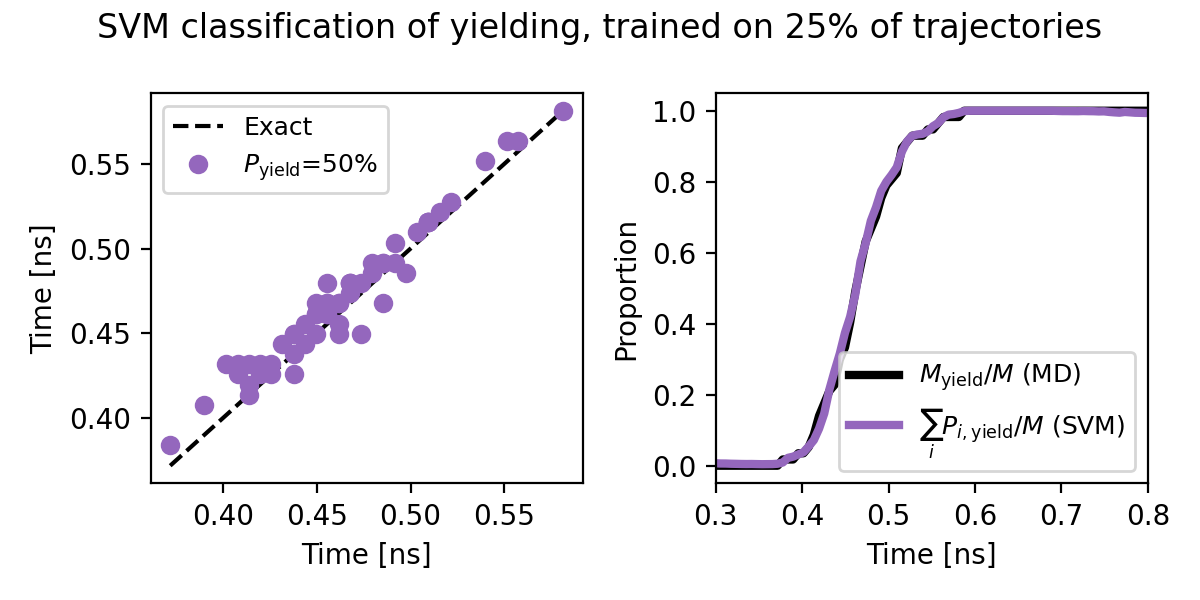}
    \caption{SVM classification of yielding. Left: scatter plot of time at yield vs predicted time at 50$\%$ probability. Right: observed and predicted proportion of yielded trajectories.}
    \label{fig:svm}
\end{figure}

\change{\subsection{Classification of yield}
A simple linear support vector machine (SVM)\cite{Rasmussen} classifier 
was trained on 25$\%$ of trajectories around the yield point. 
As can be seen in figure \ref{fig:svm}, 
the transferability to new trajectories shows excellent accuracy, 
both in pointwise comparison (left) and distributionwise (right).}

\bibliography{biblio.bib}

\end{document}